\documentclass[aps,pra,twocolumn,showpacs]{revtex4-1}
\usepackage{amsfonts}
\usepackage{amsmath}
\usepackage{amssymb}
\usepackage{xspace}

\newcommand{\prlsection}[1]{\par\medskip\emph{#1} ---}

\newcommand{\QKD}{QKD\xspace}
\newcommand{\PMQKD}{PM-QKD\xspace}
\newcommand{\EDQKD}{ED-QKD\xspace}
\newcommand{\Chaufive}{Chau05\xspace}
\newcommand{\BER}{BER\xspace}
\newcommand{\SYK}{RRDPS\xspace}
\newcommand{\LOCCtwo}{LOCC2\xspace}

\begin{document}

\title{Quantum Key Distribution Using Qudits Each Encoding One Bit Of Raw Key}
\author{H. F. Chau}
\email{hfchau@hku.hk}
\affiliation{Department of Physics and Center of Theoretical and Computational
 Physics, Pokfulam Road, Hong Kong}
\date{\today}

\begin{abstract}
 All known qudit-based prepare-and-measure quantum key distribution (\PMQKD)
 schemes are more error resilient than their qubit-based counterparts.
 Their high error resiliency comes partly from the careful encoding of multiple
 bits of signals used to generate the raw key in each transmitted qudit so that
 the same eavesdropping attempt causes a higher bit error rate (\BER) in the
 raw key.
 Here I show that highly error-tolerant \PMQKD schemes can be constructed
 simply by encoding one bit of classical information in each transmitted qudit
 in the form $(|i\rangle\pm|j\rangle)/\sqrt{2}$, where $|i\rangle$'s form an
 orthonormal basis of the $2^n$-dimensional Hilbert space.
 Moreover, I prove that these schemes can tolerate up to the theoretical
 maximum of 50\% \BER for $n\ge 2$ provided that the raw key is generated
 under a certain technical condition, making them the most error-tolerant
 \PMQKD schemes involving the transmission of unentangled finite-dimensional
 qudits to date.
 This shows the potential of processing quantum information using
 lower-dimensional quantum signals encoded in a higher-dimensional quantum
 state.
\end{abstract}

\pacs{03.67.Dd, 03.65.Aa, 89.70.-a}

\maketitle

\prlsection{Introduction}
 Quantum key distribution (\QKD) allows two cooperative players, Alice and Bob,
 to share a secret key whose security is guaranteed by the laws of quantum
 mechanics.  Since the discovery of the first \QKD scheme by Bennett and
 Brassard~\cite{BB84}, researchers have been studying different aspects of
 \QKD.  New \QKD protocols that are either more practical, efficient or error
 tolerant have been proposed.  Actual \QKD experiments for some of the
 protocols have been carried out.  Unconditionally security proofs, including
 those covering realistic settings like the use of imperfect sources and
 detectors, for many of these protocols have been found.  (See, for example,
 the review article in Ref.~\cite{Scarani09} for an overview.)

 One line of research is to investigate the use of qudits rather than qubits as
 quantum information carriers in \QKD.  In particular, Chau proved the
 unconditional security of a prepare-and-measure quantum key distribution
 (\PMQKD) scheme (called \Chaufive) using $2^n$-dimensional quantum particles
 as information carriers each encoding $n$ bits of the raw key~\cite{Chau05}.
 Although his scheme has a very low key rate and is hard to implement using
 current technology, it can tolerate a bit error rate (\BER) of up to 50\% in
 the limit of $n\to +\infty$~\footnote{There is a subtlety in defining \BER for
 qudits.  See Ref.~\cite{Chau05} for the precise definition in the case of
 \Chaufive.}.
 This demonstrates the superior error-tolerant capability of qudit-based \PMQKD
 scheme as the best qubit-based \PMQKD scheme known to date can only tolerate
 up to about 27.4\% \BER~\cite{Chau02}.  Recently, Sasaki \emph{et al.}
 proposed a radically different qudit-based \PMQKD scheme known as the
 round-robin differential-phase-shift (\SYK) protocol in which Alice encodes
 multiple bits $s_i$'s in each of the $N$-dimensional qudit state as
\begin{equation}
 \frac{1}{\sqrt{N}} \sum_{i=1}^N (-1)^{s_i} |i\rangle
 \label{E:SYK14_state}
\end{equation}
 so that Bob's measurement can only reveal one of the $(s_i - s_j)$'s of his
 choice~\cite{SYM14}.  This is a conceptually important scheme for it
 demonstrates that the security of \QKD needs not link to the Heisenberg
 uncertainty principle~\cite{Curty14}.  In terms of performance, the \SYK
 protocol can also tolerate up to 50\% \BER in the $N\to +\infty$ limit.
 Besides, if the \BER of the raw key is low, the key rate of the \SYK protocol
 is much higher than that of \Chaufive.  Several proof-of-principle experiments
 for the \SYK protocol have been conducted~\cite{Guan15,Takesue15,Wang15}.

 Here I report a family of qudit-based \PMQKD schemes whose security comes from
 a new principle.  In these schemes, Alice and Bob randomly and independently
 prepare and measure qubit-like states each in the form
 $(|i\rangle\pm|j\rangle)/\sqrt{2}$ in a $2^n$-dimensional Hilbert space for
 $n\ge 2$ so that only one bit of the raw key is encoded and transmitted in the
 phase of each qudit state.  (Here $|i\rangle$'s form an orthonormal basis of
 the $2^n$-dimensional Hilbert space.)  The security originates from the fact
 that the eavesdropper Eve has a hard time to read out a sizable portion of the
 raw key without being caught because she does not know the preparation basis
 of each qudit at the time when the quantum state is passing through the
 insecure channel under her control.  By identifying $|i\rangle$ as the single
 photon state in the $i$th optical pulse, these schemes have the additional
 attractive feature that the prepared states, which are essentially qubit
 states in diagonal basis, can be easily created and measured using a standard
 optical interferometer with variable path length.  (Interestingly, the
 experimental techniques used to prepare quantum states in
 Expression~\eqref{E:SYK14_state} in Refs.~\cite{Guan15,Takesue15,Wang15} can
 be adapted to prepare the states $(|i\rangle\pm|j\rangle)/\sqrt{2}$.)  Using
 an aggressive entanglement distillation procedure involving local operation
 and two-way classical communications (\LOCCtwo) originally reported in
 Ref.~\cite{Chau02}, I prove that Alice and Bob could share a provably secure
 secret key whenever the \BER is less than 50\% provided that the raw key obeys
 the technical condition to be stated in Eq.~\eqref{E:QKD_condition} later in
 the text, making it the first family of \PMQKD schemes that saturates the
 theoretical maximum limit of the tolerable \BER using unentangled
 finite-dimensional quantum information carriers.  This opens up the study of
 processing quantum information through the use of lower-dimensional quantum
 states embedded in a higher-dimensional Hilbert space or transferred through a
 higher-dimensional quantum channel.

\prlsection{The schemes}
 Let me denote the finite field of $N \equiv 2^n$ elements by $GF(N)$ and
 consider the following family of schemes.
\begin{description}
 \item[The family of \PMQKD schemes]
\end{description}
\begin{enumerate}
 \item Alice randomly picks $i\ne j \in GF(N)$.  She secretly prepares a state
  in the form $(|i\rangle\pm|j\rangle)/\sqrt{2}$ and sends it to Bob through an
  insecure quantum channel.
  \label{QKD:prepare}
 \item Bob randomly picks $i' \ne j' \in GF(N)$ and measures the state along
  $|i'\rangle\pm|j'\rangle$.  He keeps his measurement outcome private.
  \label{QKD:measure}
 \item By announcing the pairs $(i,j)$ and $(i',j')$ through an unjammable
  classical channel, Alice and Bob establish a bit of raw key from those states
  with $(i,j) = (i',j')$.  (They adopt the convention that $[|i\rangle + (-1)^s
  |j\rangle]/\sqrt{2}$ encodes the bit $s$.)  They discard the measurement
  outcomes of those states with $(i,j) \ne (i',j')$.  They repeat
  steps~\ref{QKD:prepare}--\ref{QKD:raw_key} until they have a long enough raw
  key.
  \label{QKD:raw_key}
 \item They estimate the \BER of the raw key $e_b$, namely, the fraction of
  mismatched bits in their shared raw bit string, by comparing (and then
  discarding) a small random sample of the raw key.  Using both accepted and
  rejected measurement outcomes in step~\ref{QKD:raw_key}, they calculate the
  conditional probability $e_c$ that a state is prepared and measured as
  $(|i\rangle\pm|j\rangle)/\sqrt{2}$ given that it is prepared as
  $(|i\rangle\pm|j\rangle) / \sqrt{2}$ and measured as $[|(1-a)i+a
  j\rangle\pm|(a+1)j-a i\rangle] / \sqrt{2}$ for some $i,j,a\in GF(N)$.  (Note
  that all arithmetic in the state-ket of a qudit is performed in the finite
  field $GF(N)$ from now on.)  They continue only if
  \begin{equation}
   e_b e_c + \frac{(N-1) (1-e_c)}{N-2} < \frac{1}{2} .
   \label{E:QKD_condition}
  \end{equation}
  \label{QKD:error_estimate}
 \item Alice and Bob apply the following \LOCCtwo classical post-processing
  procedure to the remaining raw key adapted from Ref.~\cite{Chau02}.  The
  values of the parameters $k$ and $r$ used in this procedure will be discussed
  in Methods.
  \label{QKD:priv_ampl}
  \begin{enumerate}
   \item Alice and Bob randomly group their corresponding bits in their
    remaining raw key in pairs.  They reveal the parity of each corresponding
    pair and keep the first bit in those corresponding pairs whose parities
    agree.  They repeat this process $k$~times.
   \item Alice and Bob randomly group their corresponding bits in their
    remaining raw key in sets each containing $r$ bits.  They replace each set
    by the parity of the $r$ bits in the set.
   \item Alice and Bob obtain their final secret key by applying the
    Shor-Preskill privacy amplification procedure~\cite{Chau02,SP00} to these
    bits using a Calderbank-Shor-Steane code that could correct up to, say,
    1\% quantum error.
  \end{enumerate}
\end{enumerate}

 Note that for the case of $N=4$, the above scheme takes a rather simple form.
 Alice and Bob keep those states that are prepared and measured in diagonal
 basis of the same Hilbert subspace ${\mathcal H}_{ij} \equiv \text{span}
 (|i\rangle,|j\rangle)$ for some $i\ne j\in GF(4)$.  In addition, $e_c$ equals
 the length of the raw key divided by the total number of qudits that are
 prepared in the subspace ${\mathcal H}_{ij}$ and measured in either
 ${\mathcal H}_{ij}$ or ${\mathcal H}_{ab}$ subspaces, where $i,j,a,b$ are the
 four distinct elements of $GF(4)$.

\prlsection{The unconditional security proof}
 Now I show the unconditional security~\cite{BenOr05,RK05} of this family of
 \PMQKD schemes for $N\ge 4$ by proving the unconditional security of the
 following associated family of entanglement-distillation-based quantum key
 distribution (\EDQKD) protocols using the Shor-Preskill-type
 argument~\cite{SP00}.

\begin{description}
 \item[The associated family of \EDQKD protocols]
\end{description}
\begin{enumerate}
 \item Alice prepares the state $\sum_{\ell\in GF(2)} |\ell,\ell\rangle /
  \sqrt{2}$.  She randomly picks $\lambda,\beta\in GF(N)$ with $\lambda\ne 0$
  and applies the linear transformation
  \begin{equation}
   L_{\lambda\beta}|a\rangle = |\lambda a+\beta\rangle
   \label{E:Def_L_lambda_beta}
  \end{equation}
  for all $a\in GF(N)$ to the second qudit.  She keeps the first qudit and
  sends the second qudit to Bob through an insecure quantum channel.
  \label{QKDent:prepare}
 \item Bob randomly picks $\lambda',\beta' \in GF(N)$ with $\lambda'\ne 0$ and
  applies $L_{\lambda'\beta'}^{-1}$ to the qudit he received from Alice.  Then,
  Alice and Bob projectively measure their shared state along the basis
  \begin{equation}
   {\mathcal B} = \{ |\Psi_{a\ell}\rangle \colon a\in GF(N), \ell\in GF(2) \} ,
   \label{E:Def_basis}
  \end{equation}
  where $|\Psi_{a\ell}\rangle \equiv [|0,a\rangle + (-1)^\ell |1,a+1\rangle] /
  2$.  They keep those states in the form $|\Psi_{\kappa\ell}\rangle$ with
  $\kappa\in GF(2)$ (which are regarded as qubit pairs from now on) provided
  that $\lambda = \lambda'$, $\beta = \beta'$.  They repeat
  steps~\ref{QKDent:prepare}--\ref{QKDent:measure} until they have enough
  number of shared qubits.
  \label{QKDent:measure}
 \item Let $e_{a\ell}$ be the conditional probability that the joint
  measurement outcome in step~\ref{QKDent:measure} is $|\Psi_{a\ell}\rangle$
  given that $\lambda=\lambda'$ and $\beta=\beta'$.  They continue only if
  \begin{equation}
   e_{01} + e_{11} + (N-1) (e_{10} + e_{11}) < \frac{1}{2} .
   \label{E:QKDent_condition}
  \end{equation}
  \label{QKDent:error-estimate}
 \item Alice and Bob perform the following entanglement purification
  procedure adapted from Ref.~\cite{Chau02}.
  \begin{enumerate}
   \item They randomly group their corresponding qubits in tetrads where each
    tetrad consists of two pairs shared by them.  Alice applies the unitary
    operation $|\psi_\kappa,\psi_\ell\rangle \mapsto |\psi_\kappa,\psi_{\kappa+
    \ell}\rangle$ to her share of the particles in the tetrad, where
    $|\psi_\kappa\rangle \equiv [|0\rangle + (-1)^\kappa |1\rangle]/\sqrt{2}$;
    and Bob does the same to his corresponding particles in the tetrad.  Alice
    and Bob keep their second qubit pair if the measurement results of their
    first qubit pair in the diagonal basis ${\mathcal B}^\times \equiv \{
    \psi_0,\psi_1 \}$ agree.  They repeat this process $k$~times.
    \label{QKDent:PA_phase}
   \item They randomly group their remaining qubits in sets each with $r$
    shared qubit pairs.  They separately apply the $[r,1,r]$ majority-vote
    error correction code for the rectilinear basis to their share of the
    qubits in each set.
    \label{QKDent:PA_spin}
   \item They apply a Calderbank-Shor-Steane code that could correct up to 1\%
    quantum error to the remaining shared quantum state to distill out almost
    perfect $|\Psi_{00}\rangle$ EPR pairs.  Finally, by measuring each qubit of
    these states along the diagonal basis ${\mathcal B}^\times$ Alice and Bob
    obtain their secret key.
    \label{QKDent:PA_final_key}
   \end{enumerate}
  \label{QKDent:distill}
\end{enumerate}

 Clearly $|\Psi_{a\ell}\rangle = (I \otimes {\mathtt X}_a {\mathtt Z}^\ell )
 |\Psi_{00}\rangle$ where
 \begin{equation}
  {\mathtt X}_a |b\rangle = |a+b\rangle \quad \text{and} \quad
  {\mathtt Z} |b\rangle = (-1)^{{\mathcal N}(b)} |b\rangle
  \label{E:Def_X_a_Z}
 \end{equation}
 for all $b\in GF(N)$.  Here ${\mathcal N}(b) = b^{N-1}$ is the norm of
 $b$~\cite{LN94}.  Note that $N(0) = 0$ and $N(b) = 1$ if $b\ne 0$.  Consider
 the expression
\begin{align}
 & (I \otimes L_{\lambda\beta}^{-1} {\mathtt X}_a {\mathtt Z}^\ell L_{\lambda
  \beta}) |\Psi_{b\kappa}\rangle \nonumber \\
 ={} & \sum_{\nu\in GF(2)} (-1)^{\kappa\nu + \ell {\mathcal N}(\lambda(\nu+b) +
  \beta)} |\nu,\nu+b+\lambda^{-1}a\rangle
 \label{E:L_conjugation}
\end{align}
 for all $\lambda\ne 0,\beta,a,b\in GF(N)$ and $\ell,\kappa\in GF(2)$.  Up to
 an irrelevant global phase, the R.H.S. of Eq.~\eqref{E:L_conjugation} equals
 $|\Psi_{b+\lambda^{-1}a,\kappa'}\rangle$.  Here $\kappa' = \kappa$ if $\ell=0$
 or ${\mathcal N}(\lambda b+\beta) = {\mathcal N}(\lambda(b+1)+\beta)$; and
 $\kappa'=\kappa+1$ otherwise.  Hence, the sequences of probabilities of
 measurement outcome along ${\mathcal B}$ conditioned on different
 $\lambda=\lambda'$ and $\beta=\beta'$ in step~\ref{QKDent:measure} of the
 \EDQKD protocol transform from one to another by permutation.  In addition,
 all operations in step~\ref{QKDent:distill} except the final measurement in
 the diagonal basis permute elements in ${\mathcal B}$ up to an irrelevant
 phase.  Therefore, Alice (Bob) may push the final measurement in
 ${\mathcal B}^\times$ in step~\ref{QKDent:PA_final_key} forward in time to
 immediately after step~\ref{QKDent:prepare}
 (\ref{QKDent:measure})~\cite{Chau02,LC99}.  By renaming $\lambda=j-i$ and
 $\beta=i$, I get $L_{\lambda\beta}|0\rangle = |i\rangle$ and
 $L_{\lambda\beta}|1\rangle = |j\rangle$.   Consequently, this \EDQKD protocol
 is reduced to the \PMQKD scheme.  Furthermore, the Shor-Preskill argument
 implies that the unconditional security of the above \PMQKD scheme follows
 that of the \EDQKD protocol~\cite{SP00}.

 I now proceed to analyze the security of the \EDQKD protocol.  Clearly, the
 probabilities $e_{a\ell}$'s obeys $\sum_{a\in GF(N),\ell\in GF(2)} e_{a\ell} =
 1$.  Since $\lambda$ and $\beta$ are randomly chosen for each transmitted
 qudit and are unknown to Eve during the transmission,
 Eq.~\eqref{E:L_conjugation} implies that
\begin{equation}
 e_{a0} + e_{a1} = e_{b0} + e_{b1}
 \label{E:e_ab_sum_rule}
\end{equation}
 for all non-zero $a,b\in GF(N)$.  So, if Eq.~\eqref{E:QKDent_condition} is
 satisfied, $e_{00} > 1/2$ is the greatest element among the $e_{a\ell}$'s.
 Furthermore, by comparing the definitions of $e_b$, $e_c$ in
 step~\ref{QKD:error_estimate} of the \PMQKD schemes with the definitions of
 $e_{a\ell}$'s in step~\ref{QKDent:error-estimate} of the \EDQKD protocols, I
 find the following correspondences:
\begin{equation}
 e_b e_c = (e_{01} + e_{11}) ,
 \label{E:e_b_relation}
\end{equation}
\begin{equation}
 e_c = e_{00} + e_{10} + e_{01} + e_{11}
 \label{E:e_c_relation1}
\end{equation}
 and
\begin{equation}
 1-e_c = (e_{10}+e_{11}) (N-2) .
 \label{E:e_c_relation2}
\end{equation}
 Thus, Eq.~\eqref{E:QKDent_condition} implies Eq.~\eqref{E:QKD_condition}.

 The probabilities that the joint measurement outcomes for those remaining
 shared qubits just before step~\ref{QKDent:distill} of the \EDQKD protocol can
 be written as the elements of the $2\times 2$ error matrix
\begin{equation}
 \begin{pmatrix}
  p_I & p_z \\
  p_x & p_y
 \end{pmatrix}
 \equiv
 \frac{1}{e_c}
 \begin{pmatrix}
  e_{00} & e_{01} \\ e_{10} & e_{11}
 \end{pmatrix} .
 \label{E:p_def}
\end{equation}
 By treating each pair of shared qudits as shared qubit pair, then $p_I$,
 $p_x$, $p_y$ and $p_z$ can be regarded as the probabilities that Bob's share
 of the qubit pair has suffered $I$, $\sigma_x$ $\sigma_y$ and $\sigma_z$
 errors, respectively.

 Note that in the above \EDQKD protocol, step~\ref{QKDent:distill} is analogous
 to a similar procedure in Ref.~\cite{Chau02} with the roles of $X$- and
 $Z$-errors being swapped.  That is to say, step~\ref{QKDent:PA_phase} is a
 variation of the BXOR test~\cite{B96,BDSW96} that reduces the $Z$-error of the
 resultant qubit pairs; whereas step~\ref{QKDent:PA_spin} reduces the
 $X$-error.  Applying Proposition~1 in Ref.~\cite{Chau02} with the roles of
 $X$- and $Z$-errors exchanged, the corresponding error matrix for the shared
 qubits immediately after step~\ref{QKDent:PA_phase} equals
\begin{align}
 \begin{pmatrix}
  p_I^{k\text{EP}} & p_z^{k\text{EP}} \\
  p_x^{k\text{EP}} & p_y^{k\text{EP}}
 \end{pmatrix}
 = \frac{1}{2(A+C)}
 \begin{pmatrix}
  A + B & C + D \\
  A - B & C - D
 \end{pmatrix}
 ,
 \label{E:p_ed_def}
\end{align}
 where $A = (p_I + p_x)^{2^k}$, $B = (p_I - p_x)^{2^k}$, $C = (p_y +
 p_z)^{2^k}$ and $D = (p_y - p_z)^{2^k}$.  Since $e_{00} > 1/2$, so is $p_I$.
 Hence from Proposition~2 in Ref.~\cite{Chau02} (again with $X$- and $Z$-errors
 exchanged), the quantum error rate of the shared qubits can be reduced to less
 than 1\% after step~\ref{QKDent:PA_spin} and therefore almost perfect
 $|\Psi_{00}\rangle$'s can be distilled in step~\ref{QKDent:PA_final_key} if
 the $r$ in step~\ref{QKDent:PA_spin} equals $0.005/(p_y^{k\text{EP}} +
 p_z^{k\text{EP}})$ and $2r(1/2 - p_x^{k\text{EP}} - p_y^{k\text{EP}})^2 \gg
 1$.  Such an $r$ exists if $(B+D)^2 \gg 400C(A+C)$.  Since $p_I > 1/2$, $p_I -
 p_z > p_x - p_y$.  Thus, $r$ exists by picking a sufficiently large $k$ as
 long as
\begin{equation}
 (p_I - p_x)^2 > (p_I+p_x) (p_y+p_z) .
 \label{E:condition_for_secure_key}
\end{equation}
 (Incidentally, the same condition has been proven in Ref.~\cite{Chau02} for
 the special case of $p_x = p_y = p_z$.)  From Eqs.~\eqref{E:e_b_relation}
 and~ \eqref{E:p_def} plus the fact that $e_{00} + e_b e_c + (N-1)(1-e_c)/(N-2)
 - e_{11} = 1$, the sufficient condition for the existence of $r$ can be
 rewritten as
\begin{align}
 f(e_b,e_c,e_{11}) &= \left[ 1 - e_b e_c - \frac{N(1-e_c)}{N-2} + 2 e_{11}
  \right]^2 \nonumber \\
 & \quad \quad - e_b ( 1 - e_b ) e_c^2 > 0 .
 \label{E:e_b_constraint}
\end{align}

 The maximum tolerable \BER $e_{\max}$ of the \EDQKD protocol and hence the
 \PMQKD scheme is the largest possible $e_b$ provided that the parameters $e_b,
 e_c, e_{11}$ pass the test in step~\ref{QKDent:error-estimate} of the \EDQKD
 protocol.  That is, $e_{\max} = \sup \{ e_b \colon f(e_b,e_c,e_{11}) > 0
 \text{ for all } e_c, e_{11} \text{ with } (e_b,e_c,e_{11}) \in {\mathcal R}
 \}$, where ${\mathcal R} = \{ (e_b,e_c,e_{11}) \in [0,1]^3 \colon e_b e_c +
 (N-1)(1-e_c)/(N-2) < 1/2 \}$.  Since $f$ is quadratic in $e_b$, $e_c$ and
 $e_{11}$, the value of $e_\text{max}$ can be calculated readily.
 Specifically, elements in ${\mathcal R}$ obey $1 - e_b e_c - N (1-e_c)/(N-2) >
 0$.  So, $f(e_b,e_c,e_{11}) \ge f(e_b,e_c,0)$ for all $(e_b,e_c,e_{11}) \in
 {\mathcal R}$.  Moreover, for any fixed $e_b\in [0,1/2)$ and by varying $e_c$
 in ${\mathcal R}$, it is straightforward to see that $f(e_b,e_c,0)$ is
 minimized when $e_c = e_c^*(e_b) \equiv N/[2(N-1-(N-2)e_b]$.  Finally, it is
 easy to check that $f(e_b,e_c^*(e_b),0) > 0$ if and only if $e_b \in [0,1/2)$
 provided that $N\ge 4$.  In summary, for $(e_b,e_c,e_{11}) \in {\mathcal R}$,
 $f(e_b,e_c,e_{11}) > 0$ whenever $e_b < 1/2$.  Besides, $f\to 0$ as $e_{11} =
 0$, $e_b\to 1/2$ and $e_c\to 1$.  Therefore, $e_{\max} = 1/2$; and this can be
 attained when Eve feeds every particle sent by Alice through a completely
 dephasing channel before giving it to Bob.

 By the standard composability definition of security for
 \QKD~\cite{BenOr05,RK05}, the family of \EDQKD protocols for $N\ge 4$ can,
 therefore, produce a shared secret key whenever the \BER is less then 50\%.
 
 To conclude, using the above family of \PMQKD schemes, Alice and Bob can
 establish a secure key whenever the \BER of the raw key is less than 50\%
 provided that $N\ge 4$ and the accepted data rate $e_c$ obeys
 Eq.~\eqref{E:QKD_condition}.  Since it is impossible to recover any encoded
 classical message after sending through a binary symmetric channel with
 crossover probability $1/2$, this family of \PMQKD schemes shows that the most
 error-tolerant \QKD scheme (as measured by its tolerable \BER) can be
 constructed by sending $4$-dimensional qubit-like qudits each containing a
 single bit of classical information encoded in its phase.  (The most
 error-tolerant scheme of this type using $4$-dimensional qudits before this
 study was \Chaufive, which can distill a secret key up to 35.6\% \BER.)  The
 security of this family of schemes comes partly from the ability to deduce the
 $X$-error rate through a clever use of the accepted data rate $e_c$ in
 step~\ref{QKD:error_estimate}.  This opens up new possibilities for doing
 quantum information processing through carefully designed algorithms that
 sends lower-dimensional quantum states through a higher-dimensional channel.

\prlsection{Outlook}
 So far, the analysis is restricted to the case of ideal source and detectors
 in the arbitrarily long raw key length limit.
 One still needs to investigate of the security and performance of this family
 of schemes for realistic source (say, by decoy state
 method~\cite{Wang05,LMC05,MQZL05}) and detector (say, by
 measurement-device-independent techniques~\cite{LCQ12,Tamaki14}) in the
 finite-key-length setting~\cite{Hayashi06,FMC10,HT12,CXCLTL14} using one-way
 or two-way classical post-processing.  They will be reported elsewhere.

\par\medskip
\begin{acknowledgments}
 I thank C.-H.\ F.\ Fung for his discussions, especially during the preliminary
 stage of this work.
 I also thank X.\ Ma for his discussions on the experimental implementation.
 This work is supported in part by the RGC Grant HKU8/CRF/11G of the Hong Kong
 SAR Government.
\end{acknowledgments}

\bibliographystyle{apsrev4-1}
\bibliography{qc67.4}

\end{document}